\documentclass[preprint]{aastex}

\usepackage{amssymb}
\usepackage{stmaryrd}
\usepackage{amssymb,amsmath, epsfig, epsf, lscape}

\def\beq{\begin{equation}}
\def\eeq{\end{equation}}
\def\bey{\begin{eqnarray}}
\def\eey{\end{eqnarray}}

\def\Mpc{\,{\rm Mpc}}
\def\mpc{\, h^{-1}{\rm {Mpc}}}

\def\kpc{\, h^{-1}{\rm {kpc}}}

\def\msun{\, h^{-1}{\rm M_\sun}}

\def\gs{\mathrel{\raise1.16pt\hbox{$>$}\kern-7.0pt
\lower3.06pt\hbox{{$\scriptstyle \sim$}}}}
\def\ls{\mathrel{\raise1.16pt\hbox{$<$}\kern-7.0pt
\lower3.06pt\hbox{{$\scriptstyle \sim$}}}}
\def\gtsima{$\; \buildrel > \over \sim \;$}
\def\ltsima{$\; \buildrel < \over \sim \;$}
\def\prosima{$\; \buildrel \propto \over \sim \;$}
\def\gsim{\lower.5ex\hbox{\gtsima}}
\def\lsim{\lower.5ex\hbox{\ltsima}}
\def\simpr{\lower.5ex\hbox{\prosima}}

\def\Mh{M_{\rm h}}
\def\Mth{M_{\rm th}}

\def\fof{\texttt{FoF}}

\def\mp{m_{\text{p}}}
\def\logM{\log M_h/h^{-1}M_\sun}
\def\logz{\log(1+z)}
\def\vx{\mathbf{x}}
\def\vv{\mathbf{v}}
\def\vi{\mathbf{i}}
\def\vt{\mathbf{t}}
\def\vI{\mathbf{I}}
\def\vT{\mathbf{T}}

\def\vj{\mathbf{j}}
\def\exp{\mathrm{e}}
\def\acos{\mathrm{acos}}

\slugcomment{To appear in {\it The Astrophysical Journal}}
\shorttitle{Alignments of dark matter halos} \shortauthors{Chen S. J. et al.}
\received{2016}

\begin{document}

\title{Alignments of dark matter halos with large-scale tidal fields: mass and redshift dependence}
\author{
Sijie Chen\altaffilmark{1},
Huiyuan Wang\altaffilmark{1}, H.J. Mo\altaffilmark{2},
Jingjing Shi\altaffilmark{1,3}}

\altaffiltext{1}{Key Laboratory for Research in Galaxies and
Cosmology, Department of Astronomy, University of Science and
Technology of China, Hefei, Anhui 230026, China;
souldew@mail.ustc.edu.cn, whywang@mail.ustc.edu.cn}\altaffiltext{2}{Department of
Astronomy, University of Massachusetts, Amherst MA 01003-9305,
USA; hjmo@astro.umass.edu}\altaffiltext{3}{SISSA, Via Bonomea 265, I-34136 Trieste, Italy}

\begin{abstract}
Large scale tidal field estimated directly from the distribution of dark matter halos
is used to investigate how halo shapes and spin vectors are aligned with the cosmic web.
The major, intermediate and minor axes of halos are aligned with the corresponding
tidal axes, and halo spin axes tend to be parallel with the intermediate axes and
perpendicular to the major axes of tidal field. The strengths of these alignments generally increase
with halo mass and redshift, but the dependencies are only through the peak height,
$\nu\equiv \frac{\delta_c}{\sigma(M_{h},z)}$. The scaling relations of the alignment
strengths with the value of $\nu$ indicate that the alignment strengths remain
roughly constant when the structures within which the halos reside are still
in quasi-linear regime, but decreases as nonlinear evolution becomes more important.
We also calculate the alignments in projection so that our results can be compared
directly with observations. Finally, we investigate the alignments of tidal tensors
on large scales, and use the results to understand alignments of halo pairs separated
at various distances. Our results suggest coherent structure of the tidal field
is the underlying reason for the alignments of halos and galaxies seen in
numerical simulations and in observations.
\end{abstract}


\section{Introduction}

It has been known for a while that galaxies and galaxy systems have preferred
orientations in the cosmic web.  For galaxies, their major axes  are found to have a
tendency to align with the large scale structure and with other galaxies
(Brown et al. 2002; Okumura et al. 2009; Faltenbacher et al. 2009; Zhang et al. 2013).
The spin axes of disk galaxies tend to lie in sheet-like structures (Navarro et al. 2004;
Trujillo et al. 2006; Tempel et al. 2013) and to align with the intermediate axis of the large
scale tidal field (Lee \& Erdogdu 2007; Zhang et al. 2015). For galaxy systems,
the major axes of galaxy clusters tend to point to neighboring clusters,  based on both
optical and X-ray observations (Binggeli 1982; McMillan et al. 1989; Plionis et al. 1994).
These alignments are important not only for understanding the formation and
assembly of galaxies and galaxy systems in the cosmic density field, but also
for interpreting weak gravitational lensing results,  because they may
contaminate lensing signals based on shear-shear correlations of background
sources (e.g. Croft \& Metzler 2000).

Most of the theoretical investigations so far have attempted to understand
the observed alignments through the links between galaxies and dark matter halos
extracted from cosmological $N$-body simulations. In earlier analysis, a common
practice is to assume that the spin axes of disk galaxies and the principal axes
of elliptical galaxies are directly aligned with those of their host halos
(e.g. Heavens et al. 2000; Jing 2002). The resultant alignments are,
however, much stronger than in observations, indicating that galaxies
may not be perfectly aligned with their host halos (e.g. Heymans et al. 2004;
Okumura et al. 2009). More recently, galaxies identified in hydrodynamical simulations
have been used to study galaxy alignments (e.g. \textbf{Dubois et al. 2014}; Codis et al. 2015; Velliscig et al. 2015),
and the results obtained are similar to those in observations,
indicating that baryonic processes may play an important role in producing the
observed alignments. Using cosmological simulations, these investigations
automatically take into account the coherent nature of the cosmic web
within which halos are embedded, and so galaxy alignments on large scales
may be produced by the alignments of halos with their local environments
together with the coherent structures on large scales
(e.g. Dekel et al. 1984; Splinter et al. 1997; \textbf{Faltenbacher et al. 2002}; Hopkins et al. 2005;
Schneider et al. 2012). For instance, Lee et al. (2008) measured halo
eigenvector-direction correlation function to quantify halo alignments
on large scales, and found a significant signals up to scales
of $\sim 50$ Mpc. Using the eigenvectors of the tidal tensor to represent the
direction of large scale structures, Hahn et al. (2007a,b) found that halos
have major axes preferentially parallel with the directions of the filaments
in which the halos are embedded and perpendicular to the normals of the sheets
(see also Forero-Romero et al. 2014). The spin vectors of halos
tend to be perpendicular to (parallel with) the filament and sheet
for massive (low-mass) halos (e.g. Aragon-Calvo et al. 2007; Hahn et al. 2007a,b;
Zhang et al. 2009).

In order to understand these alignment results,  Wang et al. (2011) studied the alignments
of halo spin and orientation with large scale tidal field, which is thought to be the driving force
for structure formation (Bond et al. 1996).  They found that the major and minor axes of halos
are strongly aligned with the stretching and compressing directions of the tidal field, respectively,
regardless of the morphology of the structure. Similar results were found by
Libeskind et al. (2013a) using velocity tensors. In addition, Wang et al. (2011) found
that halo spin vectors tend to be aligned with the intermediate axis and perpendicular
to the stretching direction of the tidal field (see also Forero-Romero et al. 2014),
\textbf{broadly consistent with} the tidal torque theory that works in the quasi-linear regime
(e.g. Lee \& Pen 2000; Porciani et al. 2002).

It is important to note that, in addition to halo spin and orientation, tidal field
also affects various other halo properties, such as assembly history, substructure
abundance, shape, dynamical properties and the accretion flow pattern
(e.g. Wang et al. 2011;  Shi et al. 2015; Kang \& Wang 2015). Thus,
for a given mass, halo clustering in space is expected to depend on halo
properties, a phenomenon known as halo assembly bias
(e.g. Gao et al. 2005; Wechsler et al. 2006; Wang et al. 2007; Jing et al. 2007;
Bett et al. 2007; Faltenbacher \& White 2010; Sunayama et al. 2015).
Clearly, the alignments of halos in the cosmic web provide another avenue
to investigate how environmental processes affect the formation and
structure of dark matter halos in the cosmic density field.

In this paper, we present detailed analyses of the alignments of halo orientations and spins
with the large scale tidal field, using directly the tidal tensors that can be
reconstructed from the distribution of dark halos. We focus on how
the mass and redshift dependencies of the alignments may reflect the
importance of nonlinear evolution of the cosmic density field in
affecting various alignments, and on how alignments on large and small
scales are connected to each other. The paper is organized as follows.
In section \ref{sec_sim} we describe the simulations used here, and the methods
we adopt for estimating halo principal axes, halo spins and the large scale tidal field.
In section \ref{sec_align} we present our results for halo alignments in
three dimensional space, and the dependence of alignments on redshift
and halo mass.  In section \ref{sec_pro} we show results for
the alignments in projection so that they can be compared directly
with observations. Section \ref{sec_ls} shows how tidal tensors
are aligned on large scales, and how such alignments induce
alignments of halos at large separations. Finally,  our results are
summarized in section \ref{sec_sum}.

\section{Dark matter halos and tidal fields}
\label{sec_sim}

\subsection{$N$-body simulations and Dark matter halos}

Our halo catalog is obtained from four independent cosmological
$N$-body simulations carried out with \texttt{Gadget-2} (Springel 2005).
The cosmological parameters used in these simulations
are $\Omega_{\Lambda,0}=0.742$ for the cosmological constant,
$\Omega_{\text{dm},0}=0.214$ and $\Omega_{\text{b},0}=0.044$ for
cold dark matter (CDM) and baryons,  respectively, $h=0.72$ for the dimensionless Hubble
constant, $\sigma_8=0.8$ for the {\it rms} linear mass fluctuation in a sphere of radius
$8\mpc$ extrapolated to $z=0$, and $n=1$ for the slope of the primordial fluctuation spectrum.
The CDM density field of each simulation set is traced by $1024^3$ particles in
a cubic box of $200\mpc$ on each side, with particle mass $m_\text{p}
\approx 5.3 \times 10^8 h^{-1} M_\sun$. The gravitational force is
softened isotropically on a co-moving length scale of $4\kpc$ (Plummer equivalent).
Each simulation produces 80 snapshots from $z=17$ to $z=0$, with the expansion factor evenly
spaced in logarithmic space.

Dark matter halos are selected from each snapshot with the use of the
standard \fof\ algorithm (Davis et al. 1985) with a link length equal to
0.2 times the average inter-particle separation. 
\textbf{The mass of a \fof\ halo is the sum of the masses of all particles in the halo.}
We exclude halos dominated by 'fuzzy' particles, i.e. halos whose most massive sub-halo
identified by the \texttt{SUBFIND} algorithm
(Springel et al. 2001) contains less than half of the mass of the parent \fof\ halo.
\textbf{As shown in Wang et al. (2011, see their figure 2), the `fuzzy' particle halos are more elongated and spin faster 
than normal halos, and are usually formed recently. The alignment signal for these halos can, therefore, be strongly 
affected by recent mergers. We exclude them from our analysis. 
Note that only about 4\% to 6\% of all halos with $\Mh\geq 10^{12}\msun$ are identified 
as `fuzzy' particle halos, with the fraction increasing slightly 
with increasing halo mass. Because of the small fraction of this population, excluding or including it in our analysis 
does not change our results significantly.}

\subsection{Halo principal axes and spin vector}

We use the inertia momentum tensor $\mathcal{I}$ of a halo to characterize its orientation.
The components of $\mathcal{I}$ are computed using
\beq
\mathcal{I}_{jk}=\mp \sum_{n=1}^{N} x_{n,j}x_{n,k},
\eeq
where $x_{n,j}$ ($j=1,2,3$) are the components of the position vector of the $n$th particle
relative to the center of the mass of the halo in question, and $N$ is the total number of
particles contained in the \fof\ halo. We use the normalized (unit)  eigenvectors
$\vi_1$, $\vi_2$, $\vi_3$
to denote the directions of the major, intermediate and minor axes, respectively.
The spin vector of the halo is estimated though the definition
\beq
\vj = \frac{\sum_{n=1}^{N} \vx_n \times \vv_n}{|\sum_{n=1}^{N} \vx_n \times \vv_n|}
\eeq
where $\vx_n$ and $\vv_n$ are the position and velocity vectors of the $n$th particle
relative to the center of mass, and the `$\times$' denotes cross-product.
The estimations for both the principal axes and spin vector
are affected by the mass resolution (see Schneider et al. 2012). To reduce such effects
as much as possible, we only calculate the orientation and spin vectors for halos
with masses $\Mh\geq 10^{12}\msun$, i.e. halos containing at least 1,880
particles.

\subsection{Large-scale tidal field}
\label{sec_tf}

Following Wang et al. (2011), we estimate the tidal field tensor on a halo
by summing up the tidal field tensors exerted by other halos above a mass threshold
$\Mth$.  Thus the tidal field tensor on a halo can be written as
\beq
\mathcal{T}=\sum_{n=1}^{N_h}\frac{R_n^3}{r_n^3}\mathbf{r}_n\mathbf{r}_n\,.
\eeq
Here $r_n$ ($\mathbf{r}_n$) is the co-moving distance (unit vector) from
the $n$th halos to the halo in question, $R_n$ is the virial radius of the
$n$th halo, and $N_h$ is the number of halos with masses above $\Mth$
and $r_{n}<r_{\rm p}$, with $r_{\rm p}$ being a distance limit to be specified below.
The tidal field tensor is then diagonalized to
obtain the three eigenvalues, $t_1$, $t_2$ and $t_3$
(by definition satisfy $t_1 >t_2 >t_3$ and $t_1+t_2+t_3=0$), and the
corresponding eigenvectors, $\vt_1$, $\vt_2$ and $\vt_3$ (major,
intermediate and minor axes). Defined in this way,
$\vt_1$ corresponds to the direction of stretching of the external
tidal force, while $\vt_3$ corresponds to the direction of compression.
We refer the readers to Wang et al. (2011) for the details of the tidal
field and comparisons with other environmental indicators of dark matter
halos. In the literature, another way to calculate the local tidal field
is to use the total mass density field (e.g. Hahn et al. 2007a).
In those investigations, $\mathbf{e}_3$ (or $\vt_3$) is usually used
to denote the stretching direction, while  $\mathbf{e}_1$ (or $\vt_1$)
is used  to denote the compressing direction.

In this paper we adopt $r_{\rm p}=70\mpc$. Our tests suggests that choosing
an even larger $r_{\rm p}$ changes the results very little.
In our previous studies, $\Mth=10^{12}\msun$ is adopted to
estimate the tidal field at redshift $z=0$. The co-moving number density of
halos of $\Mh\geq10^{12}\msun$ is about $2\times10^{-3}h^3\Mpc^{-3}$
at $z=0$, and so the tidal field is relatively densely sampled by
these halos. However,  the number density of such halos decreases
with increasing redshift, reaching to $\sim 10^{-5}$ at $z\sim5$,
so that only $\sim100$ such halos are available in the simulation box.
Clearly, if we want to extend the analysis to redshift $z\sim 5$,
adopting $\Mth=10^{12}\msun$ is not appropriate.

\begin{figure}
\plotone{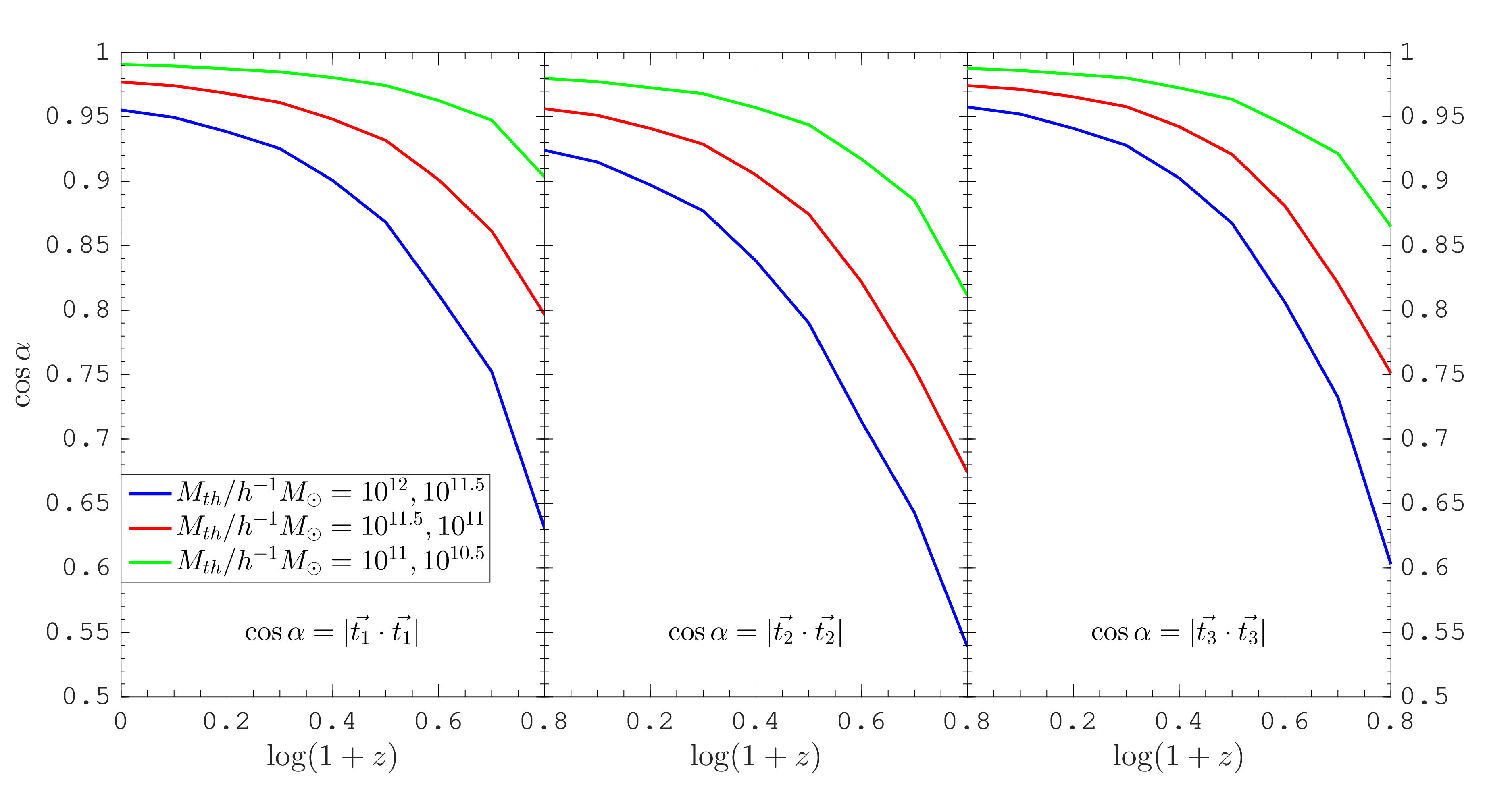}
\caption{We present the alignments between tidal fields estimated by using different halo samples as functions of redshift. The left, middle and right panels show the results for major, intermediate and minor axes, respectively. The green lines show the comparison between $\Mth=10^{10.5}$ and $10^{11}\msun$, while the blue lines show the comparison between  $\Mth=10^{11.5}$ and $10^{12}\msun$.}\label{fig_mth}
\end{figure}

To make a reasonable choice for $\Mth$, we calculate the tidal tensors
at the location of a given halo using
four different values of $\Mth$: $10^{12}$, $10^{11.5}$, $10^{11}$ and
$10^{10.5}\msun$, and estimate the difference in the orientations
of the tidal tensors  obtained from these values of $\Mth$, namely  we estimate
$\cos\alpha_{a,k}=\vert\vt_k(10^{12})\cdot\vt_k(10^{11.5})\vert$,
$\cos\alpha_{b,k}=\vert\vt_k(10^{11.5})\cdot\vt_k(10^{11})\vert$ and
$\cos\alpha_{c,k}=\vert\vt_k(10^{11})\cdot\vt_k(10^{10.5})\vert$ ($k=1,2,3$).
The mean values of $\cos\alpha$ as functions of redshift are presented in
Fig. \ref{fig_mth}. One can see that $\cos\alpha_{c,k}$ is the largest, followed by
$\cos\alpha_{b,k}$ and $\cos\alpha_{a,k}$. At $z=0$, all the three axes have
$\cos\alpha_{a,k}>0.90$, suggesting that adopting $\Mth=10^{12}\msun$ is
sufficient for a reliable estimate of the orientations of the local tidal fields.
However at $z\sim 5$, the mean $\cos\alpha_{a,k}$ decreases to
about $0.63$ for $\vt_1$, $0.54$ for $\vt_2$, and $0.60$ for $\vt_3$,
suggesting that using $\Mth=10^{12}\msun$ is no long sufficient.
On the other hand,  the mean $\cos\alpha_{c,k}$ decreases only
slowly with redshift and reaches to $0.9$, $0.81$ and $0.86$ at $z\sim5$
for the three principal axes, respectively.  It thus suggests that adopting
$\Mth=10^{11}\msun$ and $10^{10.5}\msun$ do not yield significant difference
in the estimated tidal tensor orientations even at $z\lsim5$, and that it
is unnecessary to go down to smaller  $\Mth$. In the following
presentation, all tidal fields are estimated using $\Mth=10^{10.5}\msun$.

\section{Alignments in three-dimensional space}
\label{sec_align}

The alignment of halo principal axes and spin vector with the large scale structure,
characterized either by the tidal field or velocity shear tensors, has been
investigated extensively in the past (e.g. Hahn et al. 2007b; Aragon-Calvo et al. 2007;
Zhang et al. 2009; Wang et al. 2011;  Libeskind et al. 2013a;2013b; Forero-Romero et al. 2014; \textbf{Forero-Romero \& Gonz{\'a}lez 2015}).
However, most of these investigations have been focused on $z=0$, although
some attempts have been made to extend the analysis
to moderately high redshift (e.g. Hahn et al. 2007b).
In this paper, we extend the analysis to $z\sim 5$ and study how the
alignments evolve with redshift so as to understand their origins.
We first investigate the alignments between halo principal axes and the large
scale tidal field (\S\ref{sec_hpa}) and then the alignments of halo spins
(\S\ref{sec_hs}).

\subsection{Alignments of halo principal axes with local tidal field}
\label{sec_hpa}

\begin{figure}
\plotone{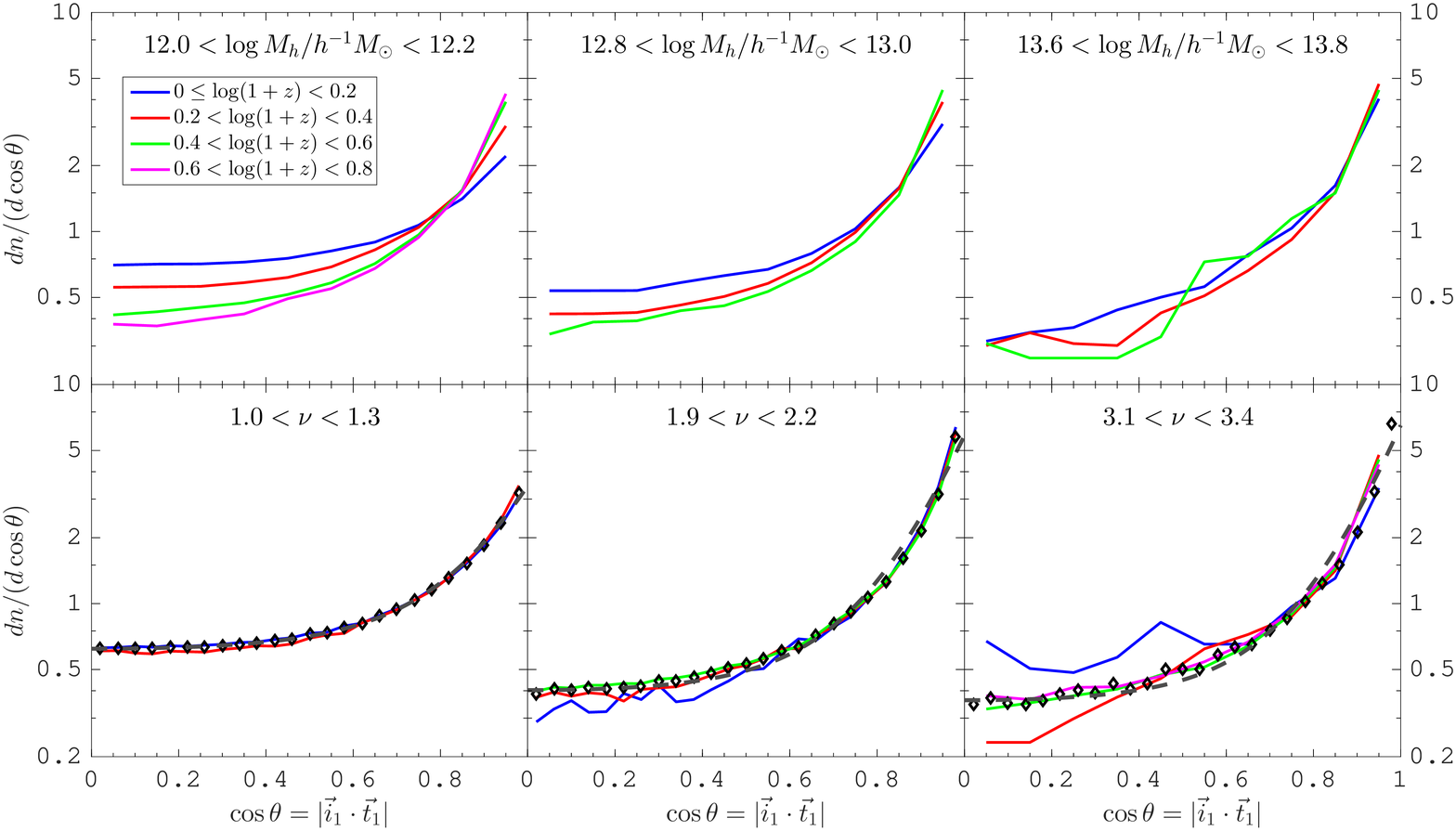}
\caption{Probability distributions of $\cos\theta=|\vi_1\cdot\vt_1|$ for halos in three mass bins(top) or three $\nu$ bins(bottom). The colored lines represent the results in four redshift ranges as indicated in the legend. The black diamonds in the bottom panels are the results averaged over all redshift range. The grey dashed lines are the fitting curves.}\label{fig_ti1d}
\end{figure}

\begin{figure}
\plotone{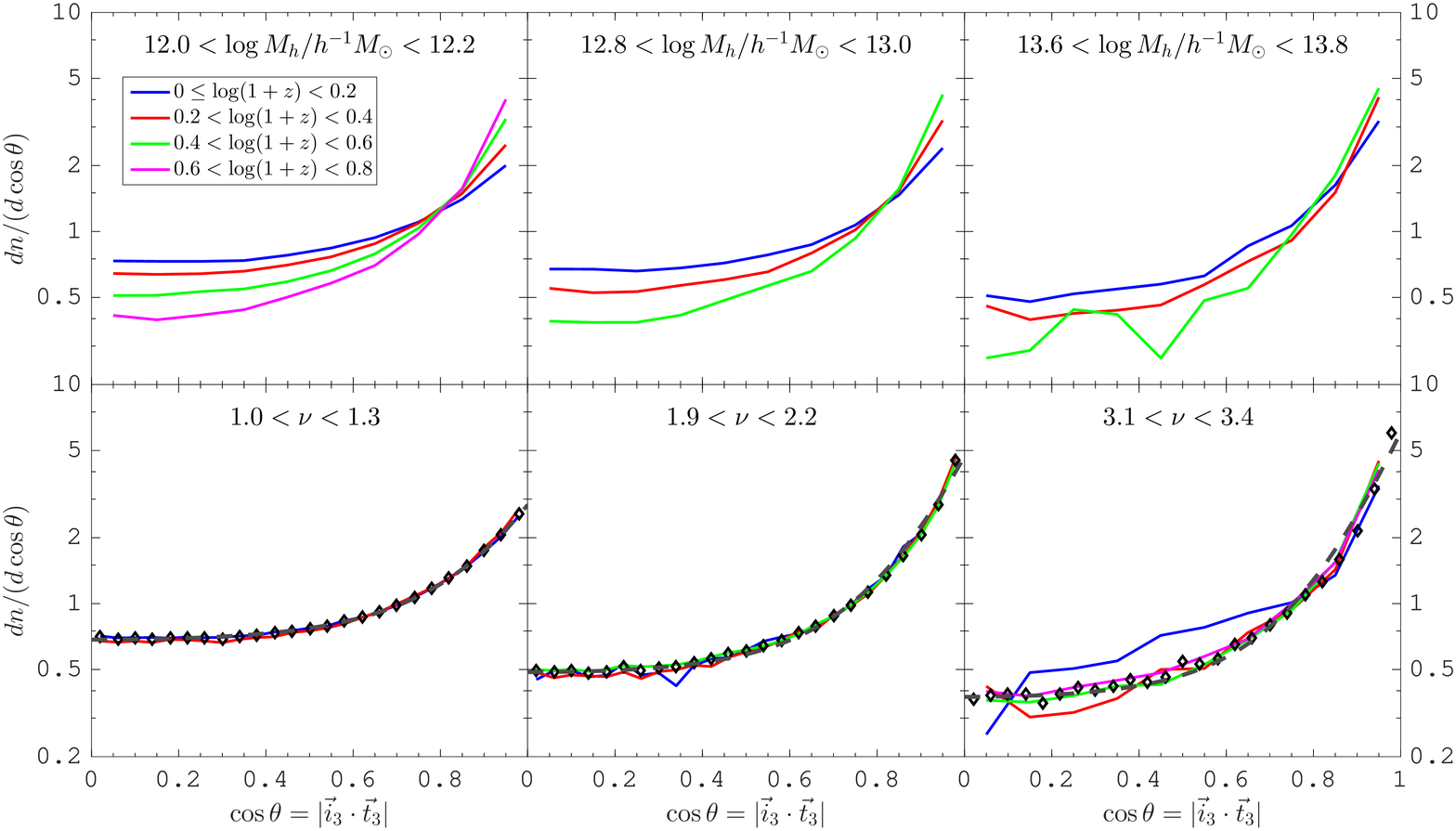}
\caption{Same as Fig. \ref{fig_ti1d} but for $\cos\theta=|\vi_3\cdot\vt_3|$}\label{fig_ti3d}
\end{figure}

Wang et al. (2011) found that the major ($\vi_1$) and minor ($\vi_3$) axes
of halos tend to align with the stretching ($\vt_1$) and compressing ($\vt_3$) directions
of the large scale tidal field, respectively (see also Libeskind et al. 2013a for similar
result based on velocity tensor). In Fig. \ref{fig_ti1d}, we show the distributions of the
cosine of the angles between $\vi_1$  and $\vt_1$ for halos in four redshift bins,
as indicated in the figure. The corresponding results for $\vi_3\cdot \vt_3$ are
presented in Fig. \ref{fig_ti3d}.  Results are shown only for $\logz\leq0.8$, as the
tidal field estimated with $\Mth=10^{10.5}\msun$ becomes unreliable at higher
$z$ (see \S\ref{sec_tf}).

The top three panels of Fig. \ref{fig_ti1d} show the results for halos in three
different mass bins, as indicated in the figure. The choices of the three mass bins
are the compromise of two considerations: first, we want to show results covering a
wide mass range; second, in each mass bin we can have at least two
relatively smooth curves to compare.  All the distributions are peaked at one,
indicating that $\vi_1$ tends to align with $\vt_1$. At a given redshift, the alignment
tends to be stronger for more massive halos, consistent with previous results
obtained for $z=0$ (e.g. Hahn et al. 2007a). For halos of the same mass, the alignment is
stronger for halos at higher $z$. However, the redshift dependence
can be almost completely eliminated if halo mass $M_h$ is expressed in terms
of the peak height $\nu$, defined by $\nu \equiv \frac{\delta_c}{\sigma(M_{h},z)}$,
where $\delta_c\approx1.686$ is the critical linear over-density for
collapse, and $\sigma(M_{h},z)$ is the rms linear mass fluctuation
on the halo mass scale extrapolated to redshift $z$.
The bottom three panels show the distributions
for halos in three $\nu$ bins selected so that for each $\nu$ bin at least two
results are reliable for comparison. Note that halos of
small masses at high redshift have $\nu$ comparable to that of the
most massive halos at $z=0$. For the highest $\nu$, the distribution
functions at the two low redshift bins are quite noisy, because the
corresponding samples contain only a small number of massive halos.
Overall our results demonstrate that the redshift dependence shown in the
top panels is produced by the evolution of the characteristic mass scale,
and that the alignment between $\vi_1$ and $\vt_1$ depends on redshift
and halo mass only through a single parameter $\nu$.

The behavior in the alignment between $\vi_3$ and  $\vt_3$ is very similar,
as shown  in Fig. \ref{fig_ti3d}. Here the dependence on redshift and halo mass
individually is even stronger than that in the $\vi_1$ - $\vt_1$ alignment, but again
the dependence is almost entirely through the peak height $\nu$.
Given that the redshift dependence is only though $\nu$, we use halos in the
whole redshift range ($\logz\leq0.8$) to obtain an overall distribution function.
The two overall functions for $\vi_1\cdot\vt_1$ and $\vi_3\cdot\vt_3$
are plotted as the black diamonds in Fig. \ref{fig_ti1d} and \ref{fig_ti3d}, respectively.
There is also a notable difference in the results between the major
and minor axes. For major axis, the alignment in the intermediate $\nu$
bin is considerably stronger than that in the smallest $\nu$ bin,
while no such difference is seen between the two higher $\nu$ bins.
For minor axis, on the other hand, the alignment continues to strengthen
with increasing $\nu$ across all the three bins. We will come back to this
difference between minor and major axes later.

\begin{figure}
\plotone{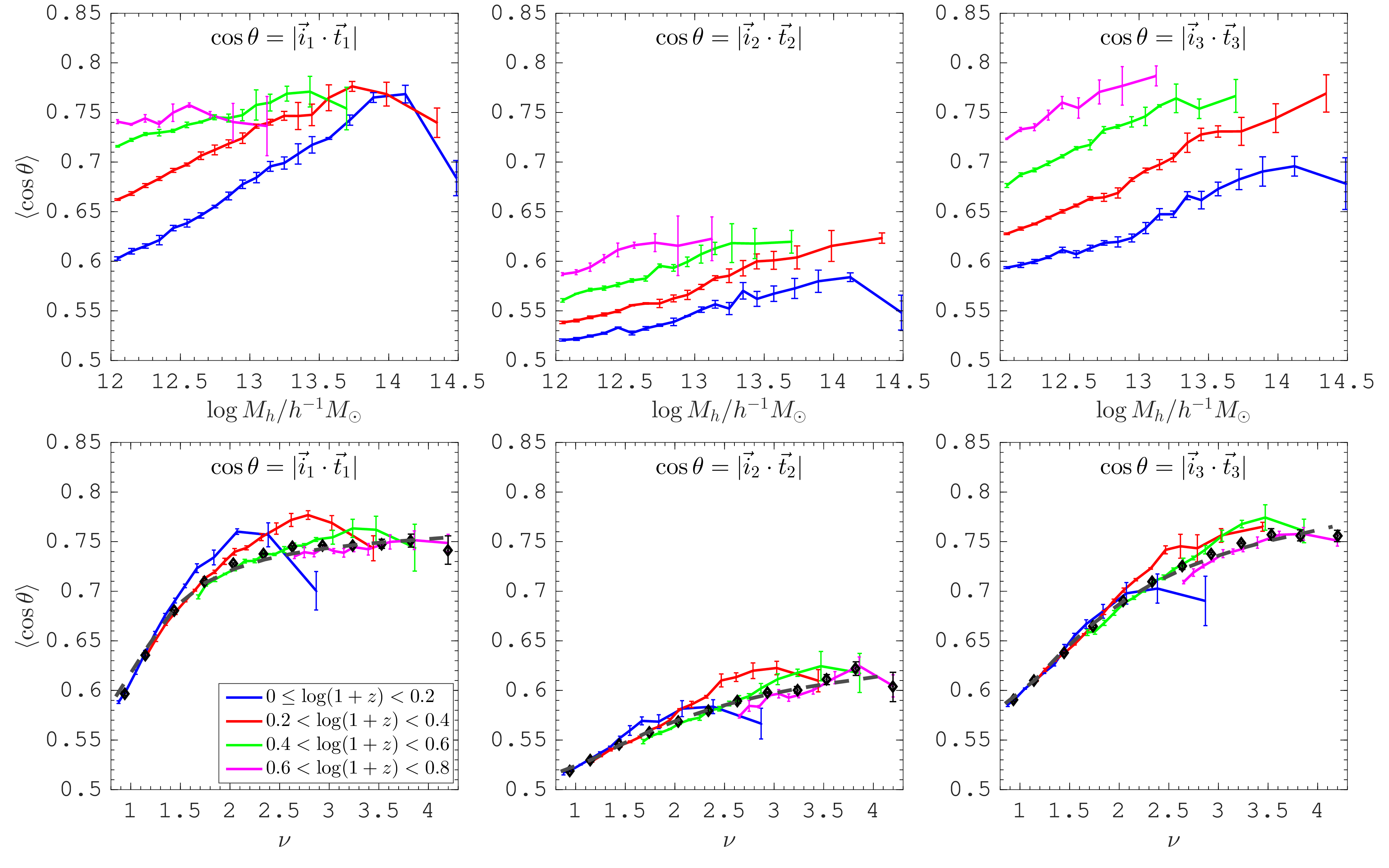}
\caption {Mean alignments as functions of $\logM$ (top) or of $\nu$ (bottom). From left to right: $\cos\theta=|\vi_k\cdot\vt_k|$, $k=1,2,3$. The colored lines represent the results at different redshift ranges as indicated in the legend and black trapezoids are the mean results averaged over all redshift. Grey dashed lines are results derived from the fitting distributions.
}\label{fig_tim}
\end{figure}

To see the dependence on redshift, halo mass and $\nu$ in more detail,
Fig. \ref{fig_tim} shows the mean cosine of the alignment angles between
the three halo principal axes and the corresponding tidal directions as
functions of halos mass and $\nu$. The four colored lines represent the
results in four different redshift bins. Bins in halo mass or in $\nu$ are equally
spaced in logarithm scale, except at the highest mass or
$\nu$ end, where wider bins are used so that
the number of halos in each bin is not too small. The error bars
(and all other error bars shown in this paper) are $1\sigma$ confidence
intervals derived from the standard deviation of the values of our
four independent simulations.

The top three panels show the alignments as functions of halo mass.
The result for $z=0$ is consistent with that obtained by Wang et al. (2011).
Overall, the strengths of the alignments increase with mass and redshift,
suggesting that the role of large scale structure in affecting halo orientation
is more important for massive halo and at high redshift.
Furthermore,  the alignments of major and minor axes have similar strength
while that of intermediate axis is weakest among the three.
For the intermediate and minor axes, the curves for the four redshift bins
share a similar positive slope. In contrast, the trends for the major
axis appear to be different: the dependence on halo mass becomes
weaker as redshift increases, and is almost absent for the highest redshift bin.

When $\nu$ is used instead of halo mass, the redshift dependence
is eliminated almost entirely for all the three axes, as demonstrated
in the bottom panels. For reference, we plot the mean values averaged
over all halos ($\logz\leq0.8$) as black diamonds.
For the major axis, the alignment strength first increases rapidly with $\nu$
and then is almost saturated above a transition scale,
$\nu_1 \simeq2.0$.  The strength of the alignment for
the minor axis increases with $\nu$ over almost the
entire range of $\nu$ that we can probe. A flat plateau also
appears, but at much higher values of $\nu$, $\nu>\nu_3\simeq3.3$.
Since the maximum values of the average cosine for both
the major and minor axes are about $0.75$, the fact that
$\nu_1<\nu_3$ implies that the alignment for the major axis is
stronger than that for the minor axis at $\nu<\nu_3$.
The alignment strength for the intermediate axis is overall much
weaker, with a maximum value of $\sim0.6$ and a transition occurring
at $\nu_2\simeq0.27$, a value between $\nu_3$ and $\nu_1$.

The results obtained here may give us some important
insights into the origin of the alignments of halos with
large-scale structure. Since $\nu\sim 1.686/\sigma(M,z)$,
and  $\sigma(M,z)$ characterizes the typical fluctuation
amplitude of perturbations of mass scale $M$ at redshift
$z$, the value of $\nu$ basically describes the importance
of non-linear environmental effects on the formation and structure
of dark matter halos, with lower $\nu$ values indicating more
important non-linear environmental effects.  The fact that
the strengths of alignments increase with $\nu$, therefore,
suggests that non-linear environments tend to weaken
the alignments.

According to our definition of tidal field, the major and minor
axes correspond to the stretching and compressing directions of the
local gravitational field. Thus, gravitational collapse to form a halo
is expected to proceed from being along the minor axis first,
then along the intermediate axis, and lastly along the major
axis of the local tidal field. As such, non-linear evolution is expected
to be the most important along the minor axis and the least along
the major axis. Consequently, for a given $M$, non-linear effects, which
tend to suppress alignment, start to operate earlier along the minor axis,
i.e. when $\sigma(M,z)$ is smaller or $\nu$ is larger, than along
the other two axes. This explains why $\nu_1<\nu_2<\nu_3$.
According to this interpretation, $\nu_k (k=1,2,3)$ may be used to
indicate the transition of the environmental effects from
the linear to nonlinear regimes.  While nonlinear effects become
important to affect the alignments at  $\nu<\nu_k$,  the results
at $\nu>\nu_k$ mainly reflect the alignments between halos with the
linear tidal field. Our results, therefore,  show that, in the linear regime,
the alignments between halos and tidal tensor are quite independent
of $\nu$. Similar behavior is also found in the alignments of halo
spins with tidal tensor,  as we will see in the next subsection.

Another possibility is that non-linear processes do not play any
important role in affecting the alignments of halos with tidal tensor,
and  the dependence on $\nu$ is completely due to the
initial alignments in the linear density field. The dependence of
the alignment strength on $\nu$ may then be explained by the fact that
halos of different $\nu$ reside in different local tidal fields.
However, it is unclear how this scenario explains the difference
in the transition scales for the three different axes.

Using the tidal field estimated from the mass density field,
Hahn et al. (2007b) found that the major axes of halos embedded in
filamentary structures tend to be parallel with the filament,
while the major axes of halos in sheets are preferentially parallel to
the sheet plane. In particular, they found that these alignments are
independent of redshift once the halo mass is scaled with the typical halo
mass, $M_*$,  defined through $\delta_c/\sigma(M_*,z)=1$.
Our findings are consistent with theirs,  but there are several important differences.
In the investigation of Hahn et al. (2007b), only halos with
$z\leq1$ are considered, while our analyses extend to much higher
redshift, $z\sim5$. As we have demonstrated, including halos at high
redshifts is crucial in revealing the regime where the dependence
on $\nu$ becomes unimportant. When presenting their
results, Hahn et al. adopted $2M_{\ast}$ as
the smoothing scale to calculate the tidal tensor.  Since $M_{\ast}$
decreases rapidly with increasing redshift, the smoothing scale
will become too small at high redshift to be defined properly in
simulations. For instance, at $z=5$, $\log M_{\ast}\simeq6.3$ for
the WMAP5 cosmology, and so the smoothing mass scale of
$2M_{\ast}$ corresponds to a length scale of  $0.015\mpc$,
which is usually much smaller  than the grid size used in calculating
the tidal tensor. This might be the reason why Hahn et al. did not
go beyond $z=1$. In our analysis, the tidal field is estimated from
the halo population, and our tests have shown that the method
provides a reliable estimate of the tidal tensor at $z=5$
when halos with masses down to $\Mth=10^{10.5}\msun$ are
used. In addition, instead of using the large-scale structures, such
as filaments and sheets to represent the large scale environments
of halos,  as was done in Hahn et al., we use directly the local tidal tensor
that is more closely related to accretion patterns around
dark matter halos (see Shi et al. 2015). Finally, the tidal field derived from
the mass density field includes the contribution of halo's self-gravity.
This led Hahn et al.  to suggest that the dependence of the alignment
strength on halo mass is due to the fact that their tidal tensor estimate
may be affected by halo shapes, which are more elongated for more
massive halos. Our estimate of the tidal field does not include
the self gravity of halos, and so our results are not affected by the
mass-dependence of halo shape.

\subsection{Halo spins}
\label{sec_hs}

The tidal torque theory predicts that the halo spin axis tends to be parallel with
the intermediate axis of the tidal field, i.e. with $\vt_2$
(e.g Lee \& Pen 2000; Porciani et al. 2002; Lee \& Erdogdu 2007).
To test this with our simulations, we show in Fig. \ref{fig_jt2d}
the distributions of $\cos\theta=|\vj\cdot\vt_2|$ for the same three mass bins
and four redshift bins as used above for the halo principal axes.
As one can see, the halo spin axis tend to align with $\vt_2$,
but the trend is not strong,  with the alignment strength
increasing with halo mass. These results are in good agreement with
previous findings, and provide support to the tidal torque theory.
In addition, our results also reveal that, for a given halo mass,
the alignment of the spin axis with $\vt_2$ tends to be stronger at higher redshift.

Here again, the dependence on redshift and halo mass is through
the peak height, $\nu$,  and the redshift dependence is almost
entirely eliminated when $\nu$ is used instead of halo mass, as shown
in the lower panels. Note that for the highest $\nu$ bin, the two high redshift
curves match each other very well; the discrepancy seen for the two lower
redshift curves is mainly caused by small number statistics, as halos
with high $\nu$ are rare at low $z$. The black diamonds in the
lower panels  show the results obtained by averaging over halos
in the entire redshift range.

Fig. \ref{fig_jtm} shows the mean values of the cosine of the alignment angles between
$\vj$ and $\vt_k(k=1,2,3)$ as functions of halo mass (top panels).
In contrast to $\vt_2$, $\vt_1$ tends to be perpendicular to $\vj$
(see e.g. Wang et al. 2011; Forero-Romero et al. 2014). The strengths
of both the alignment with $\vt_2$ and the anti-alignment with $\vt_1$
increase with increasing redshift and halo mass. The $\vj$-$\vt_3$ alignment
appears more complicated. First, the dependence of the alignment
strength on halo mass and redshift is weaker than those for the
other two axes, $\vt_1$ and $\vt_2$. Second, massive halos tend to have
their spin direction weakly aligned with $\vt_3$, while
the ones with $\logM<13.5$ exhibit a weak but significant anti-alignment
that  is almost independent of $z$ at $z<3$.

The bottom three panels of Fig. \ref{fig_jtm} show the mean alignment angles
as functions of $\nu$ (instead of halo mass). It is remarkable that the redshift
dependence seen in the upper panel for the intermediate axis is almost entirely
eliminated.  The overall mean values together with the fitting results for the
spin alignment with intermediate axis are also plotted in the lower middle
panel for reference. One sees that the strength of the alignment first increases
with increasing $\nu$, and then remains roughly at a constant value of
$\sim0.57$ at $\nu>\nu_j\simeq 2.5$. This behavior is very similar to
that seen in the alignments of halo principal axes with the tidal tensor.
In particular, the transition scale, $\nu_j$, for the spin alignment
is very close to $\nu_2$ in the $\vi_2$-$\vt_2$ alignment,
indicating that the two alignments may have a similar origin.

Based on the tidal torque theory, Porciani et al. (2002) showed that
the mean cosine of the alignment angle between $\vj$ and $\vt_2$ is
$0.59$ for halos more massive than $10^{12}\msun$. They also calculated
such alignment for proto-halos and obtained a value of $0.56$. These results
are in good agreement with ours ($0.57$) for high $\nu$.
At lower $\nu$, our simulations give lower alignment strengths
than the theoretical predictions. This may not be surprising,  because
a lower value of $\nu$ implies that non-linear effects are more
important (see \S\ref{sec_hpa}) and because the tidal torque
theory is expected to work well only in the quasi-linear regime.
Thus, our results suggest that the strength of the spin-$\vt_2$ alignment
in the linear regime is, on average, a constant over a large mass range,
and non-linear effects tend to reduce the alignment.

The situations for the other two axes are more complicated. For halos of a given
mass, redshift dependence in the strengths of the $\vj$-$\vt_1$ and $\vj$-$\vt_3$
alignments is clearly present, and particularly strong in the former,
as shown in the upper panels of Fig. \ref{fig_jtm}. The use of $\nu$
to replace halo mass shifts the results for the high redshift bins
to the right, but the shift is so much that a reversed redshift
trend is produced. Compared to the results shown in the
top panels, the four lines for $\vj$-$\vt_1$ are now closer,
particularly at small $\nu$, although still not on top of each other.
Since halos acquire their angular momenta through accretion,
the correlation of spin vector with the major and minor axes of tidal field
may be understood in terms of accretion flow.  In Shi et al. (2015) it
was found that the position and velocity vectors of the accreted sub-halos
relative to the hosts tend to be parallel with, and perpendicular to,  the major
axis of the tidal field, respectively. Thus, the accreted angular momentum is
expected to be preferentially perpendicular to the major axis, as we
see in the results for the $\vj$-$\vt_1$ alignment. They also found that
the position vector tends to be perpendicular to the minor axis,
but the alignment between velocity vectors and the minor axis is weak.
This is consistent with the weak $\vj$-$\vt_3$ alignment we see here.
However, it is still unclear why  the $\vj$-$\vt_1$ and $\vj$-$\vt_3$
alignments do not have as tight a scaling relation with $\nu$ as
the $\vj$-$\vt_2$ alignment does.

\subsection{Fitting to the scaling relations}

Our results above show that the alignments between
$\vi_1$ and $\vt_1$, $\vi_2$ and $\vt_2$, $\vi_3$ and $\vt_3$,
and $\vj$ and $\vt_2$ all obey some scaling relations with the
peak height $\nu$. In this subsection, we present the fitting
results for these relations. As shown above, these relations
all seem to contain two phases, and we adopt the following form
to fit the mean alignments as functions of $\nu$,
\beq\label{fiteq1}
A(\nu)=a_1+a_2{\rm arctan}(a_3(\nu-0.85))\,.
\eeq
The fitting results are shown in Figure \ref{fig_tim}
and \ref{fig_jtm} as dashed lines. Given the uncertainties, the
fitting results describe the simulation data well.
The best fitting parameters are $(a_1,a_2,a_3)=(0.57, 0.13, 2.06)$
for $\vi_1$-$\vt_1$, (0.51, 0.09, 0.61) for $\vi_2$-$\vt_2$,
(0.58, 0.16, 0.73) for $\vi_3$-$\vt_3$ and (0.58, 0.16, 0.73) for $\vj$-$\vt_2$.

\begin{figure}
\plotone{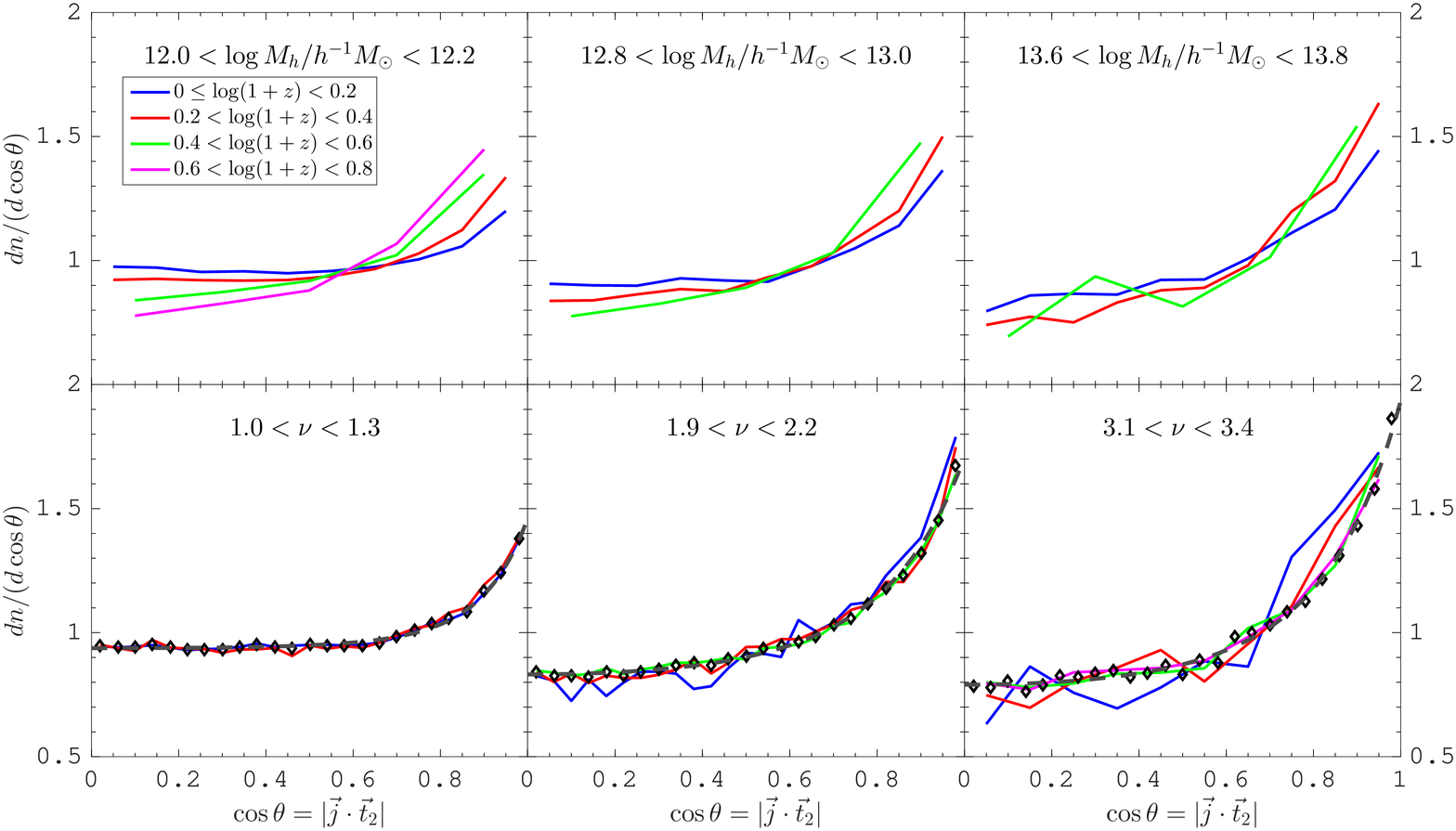}
\caption {Probability distributions of $\cos\theta=|\vj\cdot\vt_2|$ for halos in three mass bins(top) or three $\nu$ bins(bottom). The colored lines represent the results in four redshift ranges as indicated in the legend. The black diamonds in the bottom panels are the results averaged over all redshift range. The grey dashed lines are the fitting curves.
}\label{fig_jt2d}
\end{figure}

\begin{figure}
\plotone{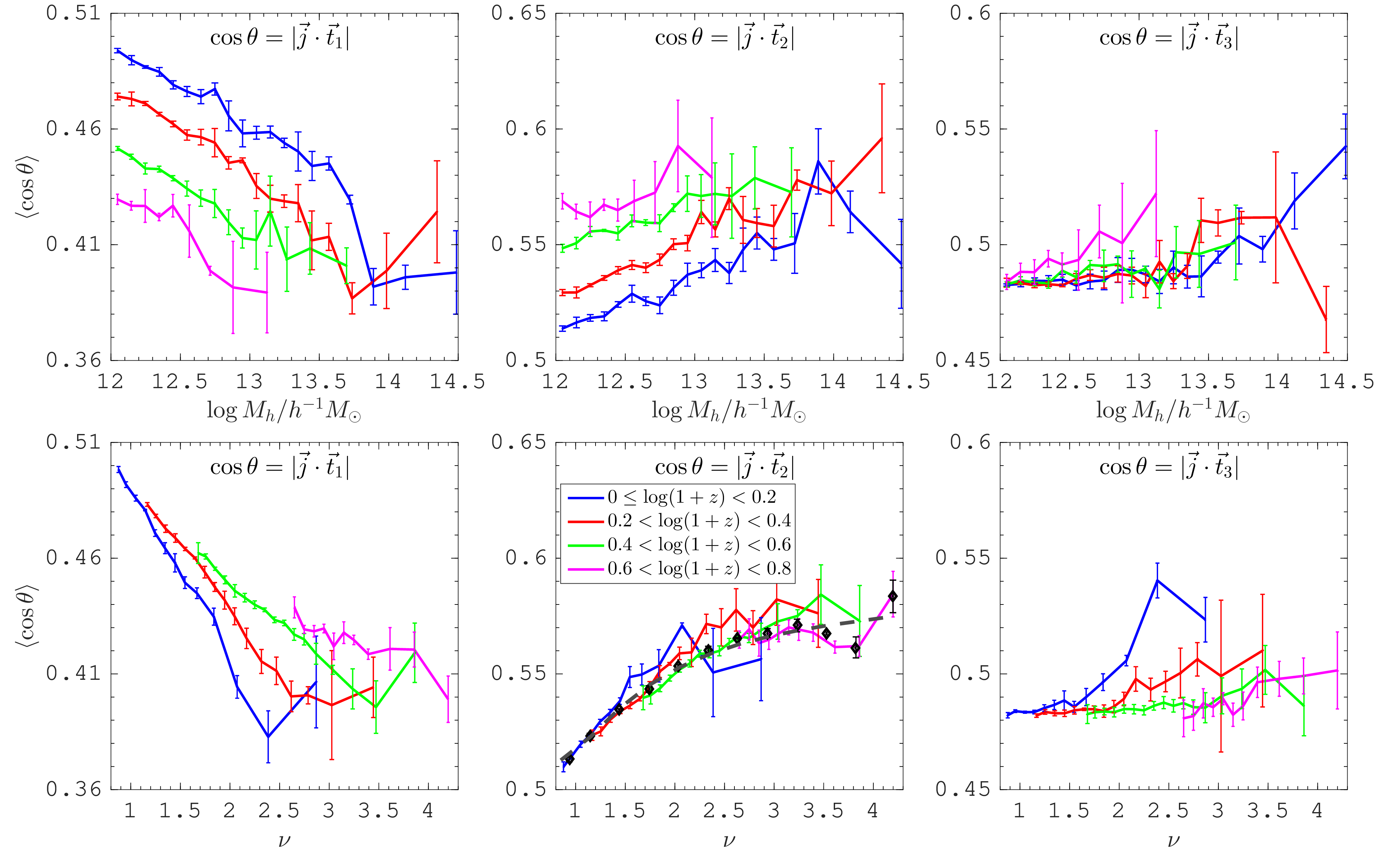}
\caption {Mean alignments as functions of $\logM$ (top) or of $\nu$ (bottom). From left to right: $\cos\theta=\vj\cdot\vt_k$, $k=1,2,3$. The colored lines represent the results at different redshift ranges as indicated in the legend and black trapezoids are the mean results averaged over all redshift. Grey dashed lines are results derived from the fitting distributions. }\label{fig_jtm}
\end{figure}

For each $\nu$ bin, we find that the overall distribution of $\cos\theta$
(shown as diamonds in Figure \ref{fig_ti1d}, \ref{fig_ti3d} and \ref{fig_jt2d})
can be well described by the following form
\beq\label{fiteq2}
\frac{d\,n}{d\cos\theta}=\frac{2}{1-2\beta (\exp^b-1)}
\left\{ A(\nu) - \beta\left(\exp^b-1\right)+\beta\left[1-2A(\nu)\right] b\exp^{b\cos^2\theta}\right\}\,,
\eeq
where $b$ is the only free parameter,
$\beta=[\sqrt{\pi b} \mathrm{erfi}(\sqrt{b})]^{-1}$ and $\mathrm{erfi}$ is the imaginary error function.
The form of this equation makes sure that the integral from $\cos\theta=0$ to $1$ is
equal to one (i.e. the distribution function is normalized) and the mean
$\cos\theta$ of the distribution gives $A(\nu)$.
The Levenberg-Marquardt method is used to find the
best fitting parameter $b$, and the results for different $\nu$
bins are given in Table 1. The corresponding curves are plotted in
Figures \ref{fig_ti1d}, \ref{fig_ti3d}  and  \ref{fig_jt2d} as the
dashed lines for the three $\nu$ bins we have chosen to plot.
Note that the $\vi_2$-$\vt_2$ alignment have properties similar
to the other two axes. The fitting parameters for this axis are
also given in the table for completeness, although the alignment
results are not shown in figures.

\begin{table}
    \begin{center}
    \caption{The fitting parameters of Eq. \ref{fiteq2} for the alignments}
\begin{tabular}[c]{|c|c|c|c|c|}
    \hline
    \hline
    $\nu$ bin	    &$|\vi_1\cdot\vt_1|$
					&$|\vi_2\cdot\vt_2|$
					&$|\vi_3\cdot\vt_3|$
					&$|\vj\cdot\vt_2|$  \\
	\hline
	$0.7\sim1.0$	&$3.80$  &$3.71$  &$3.28$  &$5.70$	\\	
	$1.0\sim1.3$	&$4.04$  &$3.45$  &$3.59$  &$4.60$	\\	
	$1.3\sim1.6$	&$4.47$  &$3.23$  &$3.90$  &$3.57$	\\	
	$1.6\sim1.9$	&$4.79$  &$3.19$  &$4.23$  &$2.97$	\\	
	$1.9\sim2.2$	&$5.01$  &$3.07$  &$4.48$  &$2.44$	\\	
	$2.2\sim2.5$	&$5.04$  &$2.85$  &$4.77$  &$2.03$	\\	
	$2.5\sim2.8$	&$5.19$  &$2.85$  &$4.86$  &$2.11$	\\	
	$2.8\sim3.1$	&$5.29$  &$2.76$  &$5.00$  &$2.65$	\\	
	$3.1\sim3.4$	&$5.23$  &$2.89$  &$5.05$  &$1.94$	\\	
	$3.4\sim3.7$	&$4.62$  &$2.55$  &$5.22$  &$1.86$	\\	
	$3.7\sim4.0$	&$4.69$  &$2.20$  &$5.22$  &$1.19$	\\	
	$4.0\sim4.3$	&$4.79$  &$3.01$  &$5.91$  &$0.10$	\\	
    \hline
    \hline
	\end{tabular}
    \end{center}
	\end{table}

\section{Alignments in projection}
\label{sec_pro}

In observation, three-dimensional tidal field can be reconstructed
from the distribution of galaxy groups (e.g. Wang et al. 2012) or
from galaxy distribution  (e.g. Lee \& Erdogdu 2007). However,
halo orientations in 3-d space are difficult to obtain observationally.
One common practice is to study the alignments between large scale
structure and the following two projected orientations:
(i) the projected distribution of satellite galaxies; (ii) the orientation of
the image of the central galaxy in a group. In this subsection we present
alignment results in two-dimensional space, which may be more closely
related to observation.

We choose the $x$-$y$ plane in a simulation to represent the
sky, namely the $z$-axis to be along the line of sight.
For a halo, the projected principal axes are then represented by the
following vectors in the $x$-$y$ plane:
\beq\label{vIdef}
\vI_k=\frac{1}{\sqrt{i_{k,1}^2+i_{k,2}^2}}(i_{k,1},i_{k,2})~~ (k=1,2,3)\,,
\eeq
where $i_{k,1}$ and $i_{k,2}$ are the $x$ and $y$ components of the three
dimensional principal axis, $\vi_k$. Similarly, the projected tidal
directions are given by
\beq\label{vTdef}
\vT_k=\frac{1}{\sqrt{t_{k,1}^2+t_{k,2}^2}}(t_{k,1},t_{k,2})~~(k=1,2,3)\,,
\eeq
where $t_{k,1}$ and $t_{k,2}$ are the $x$ and $y$ components of $\vt_k$.

\begin{figure}
\plotone{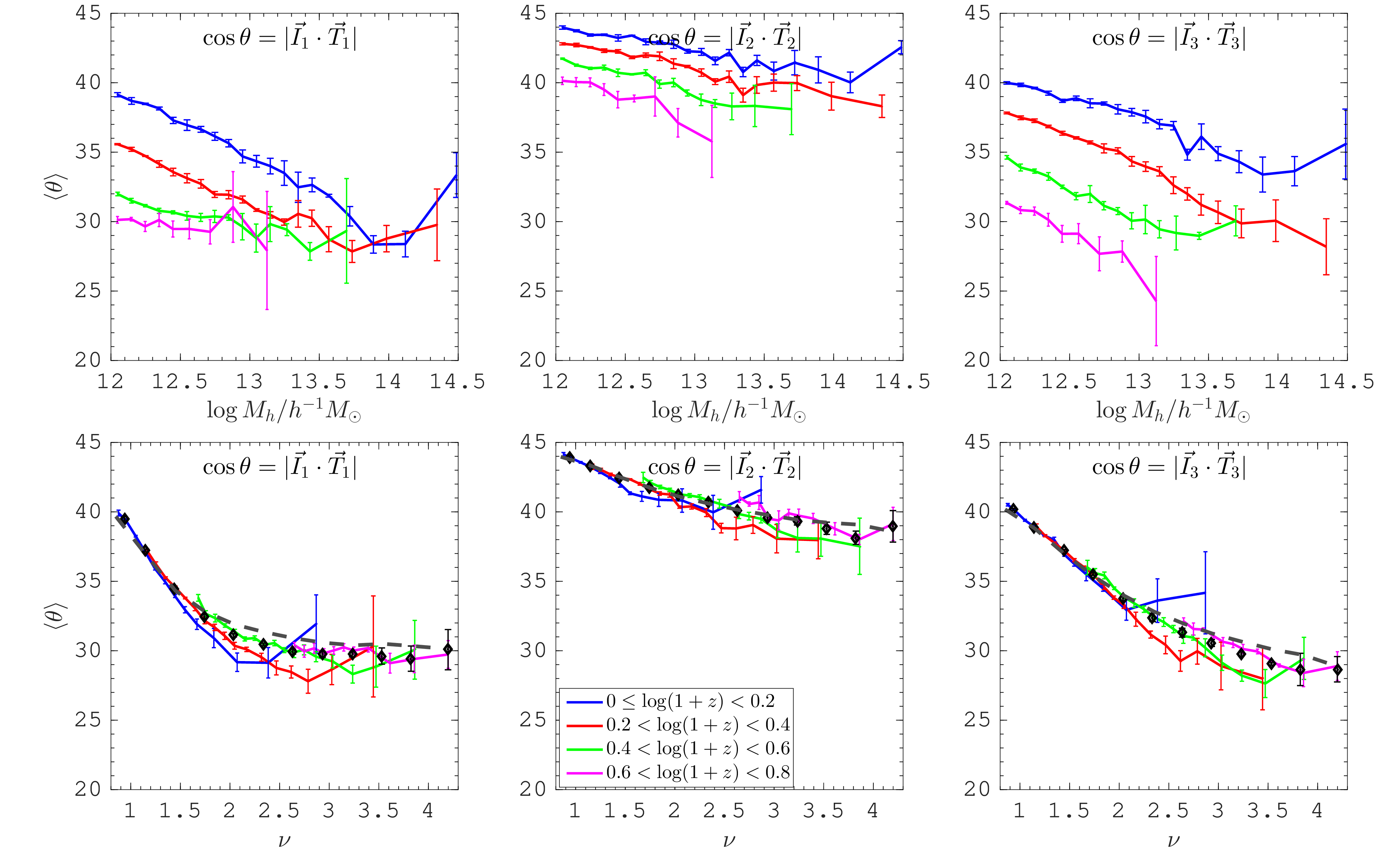}
\caption {Mean alignment angle $\theta$ as functions of $\logM$ (top) or $\nu$ (bottom) for projected vectors. From left to right: $\theta=\acos(|\vI_k\cdot\vT_k|)$~$k=1,2,3$, where $\vI_k$ and $\vT_k$ are projected principle axes of halo and tidal field, as defined in \eqref{vIdef} and \eqref{vTdef}. The colored lines represent the results at different redshift ranges as indicated in the legend. }\label{fig_ptim}
\end{figure}

The mean alignment angles between $\vI_k$ and $\vT_k$ as functions of halo mass
and $\nu$ are plotted in Fig. \ref{fig_ptim}. Here we show the alignment angles instead of the
their cosines. In two-dimensional space, the angle between two random
vectors has a uniform distribution in the alignment angle, while for two random vectors
in the three-dimensional space it is the cosine of the angle between the two vectors
that has a uniform distribution. As is clear, there is a strong tendency for $\vI_k$ to be
aligned with $\vT_k$, as the average angles are all smaller than the expected
value of $45^\circ$.  The alignment is stronger  for halos of higher masses and
and at higher redshift. The dependence on mass and redshift is largely through
$\nu$, as shown in the lower panels. The diamonds in the
lower panels show the mean angles obtained from the entire redshift range,
$\log(1+z)<0.8$, and the dashed curves are derived from the fitting results
in the three-dimensional case.

Our simulation results agree qualitatively with the observational results based on the
orientations of  central galaxies (see e.g. Zhang et al. 2013). However, the alignments
obtained here are much stronger than that based on central galaxies, as is expected
because central galaxies are not perfectly aligned with their host halos. Indeed,
as shown in Kang et al. (2007),  in order to reproduce the alignment between the satellite
distribution and the central galaxy orientation,  central galaxies have to have certain
misalignment with their host halos (see also Wang et al. 2008). Moreover,
as shown in Shi et al. (2015), the inner part of a halo,  which may be more
relevant to the properties of central galaxy (e.g. Wang et al. 2014b), is less
strongly correlated with tidal field than the outer part.

In observations, the spin axis of a spiral galaxy is usually obtained from the axis
ratio of its image, assuming that the disk is intrinsically round and that the spin
axis is perpendicular to the disk. Since it is usually unknown which side of the
disk is closer, namely the sign of the $z$ component of the spin vector is not
determined,  a given axis ratio corresponds to two vectors with
opposite signs for the $z$-component (e.g. Tempel et al. 2013).  To account
for such uncertainty, we define two spin directions
for each halo, $\vj_+\equiv\vj=(j_1,j_2,+j_3)$ and $\vj_-\equiv(j_1,j_2, -j_3)$,
where $j_l$ ($l=1,2,3$) are the three components of the halo spin vector.
The observed alignment should be the mean values averaged over the
alignments of the two vectors.

\begin{figure}
\plotone{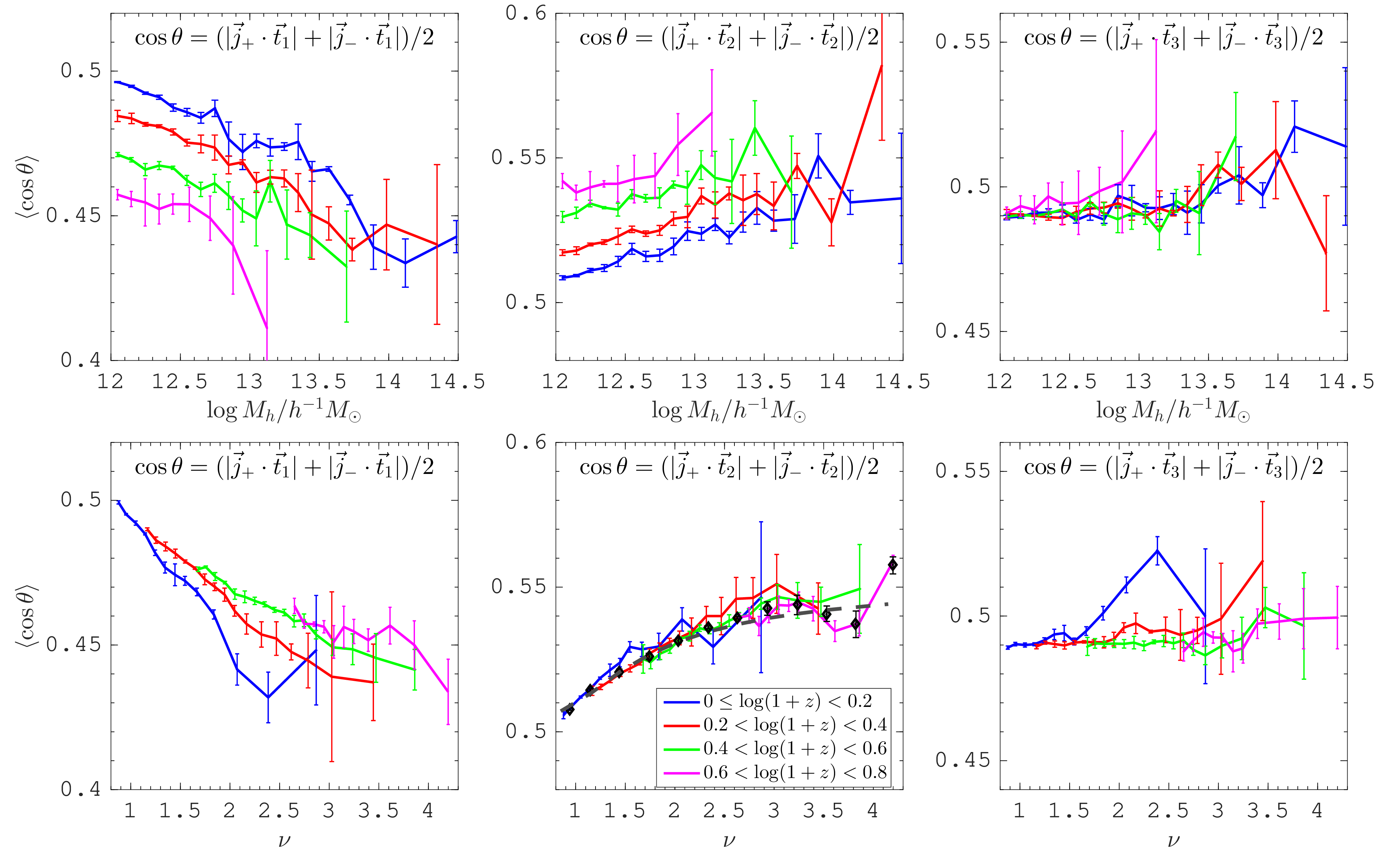}
\caption {Mean alignments as functions of $\logM$ (top) or $\nu$ (bottom). From left to right: $\cos\theta_k=(\cos\theta_{k,+}+\cos\theta_{k,-})/2$, where $\cos\theta_{k,+}=|\vj_+\cdot\vt_k|$ and $\cos\theta_{k,-}=|\vj_-\cdot\vt_k|$, $k=1,2,3$. Here $\vj_{\pm}$ are artificial spin axes, which are used to account for the projection effect.
The colored lines represent the results at different redshift ranges.}\label{fig_ptjm}
\end{figure}

Fig. \ref{fig_ptjm} shows the mean values between
$\langle |\vj_-\cdot\vt_k|\rangle$ and $\langle |\vj_+\cdot\vt_k|\rangle$ as functions of
halo mass and $\nu$. The spin vectors defined in this way still tend to be parallel
with $\vt_2$ and perpendicular to $\vt_1$, but the strengths of the alignments
are reduced in comparison to the full three dimensional cases, as is expected.
As in the full three-dimensional case, for a given halo mass, the alignment
of the projected spin with $\vt_2$ depends strongly on redshift, but the dependence
can be eliminated if $\nu$ is used instead of halo mass.

Recently, Zhang et al. (2015) performed similar analyses
and found $\langle |\vj_{\pm}\cdot\vt_{k=1,2,3}|\rangle =0.500, 0.508, 0.488$
for halos of masses $\sim 10^{12}\msun$ and $0.477, 0.519, 0.497$ for halos of masses
$\sim10^{13}\msun$, in good agreement with our results.
\footnote{Note that they used $\vt_3$ to denote the stretching direction and
$\vt_1$ to denote the compressing direction.} As shown
in Zhang et al. (2015), the predicted halo spin - tidal tensor alignments
are stronger than the observed results derived from disk galaxies,
but the disagreement can be mitigated if spins of the inner parts of halos
are used in the model predictions.

\section{Alignments on large scales}
\label{sec_ls}

Alignments of halo orientations on large scales (e.g. Hopkins et al. 2005;
Lee et al. 2007) are important to understand, because such alignments
may produce galaxy-galaxy alignments on large scale, thereby affecting
the interpretations of gravitational lensing results (e.g. Heavens et al. 2000;
Jing 2002; Heymans et al. 2004). As described above, halos show strong alignments
with their local tidal tensors (halo-tidal tensor alignments). If the
tidal tensors at locations separated by large distances are aligned too
(tide - tide alignments), then halo-halo alignments on large scales may be
understood as a result of these two kinds of alignments. In this section
we investigate these large-scale alignments, first  (in \S\ref{sec_ctd})
focusing on the tide-tide alignments, and then (in \S\ref{sec_cho})
on halo-halo alignments on large scales.

\subsection{Alignments of tidal tensors on large scales}
\label{sec_ctd}

\begin{figure}
\plotone{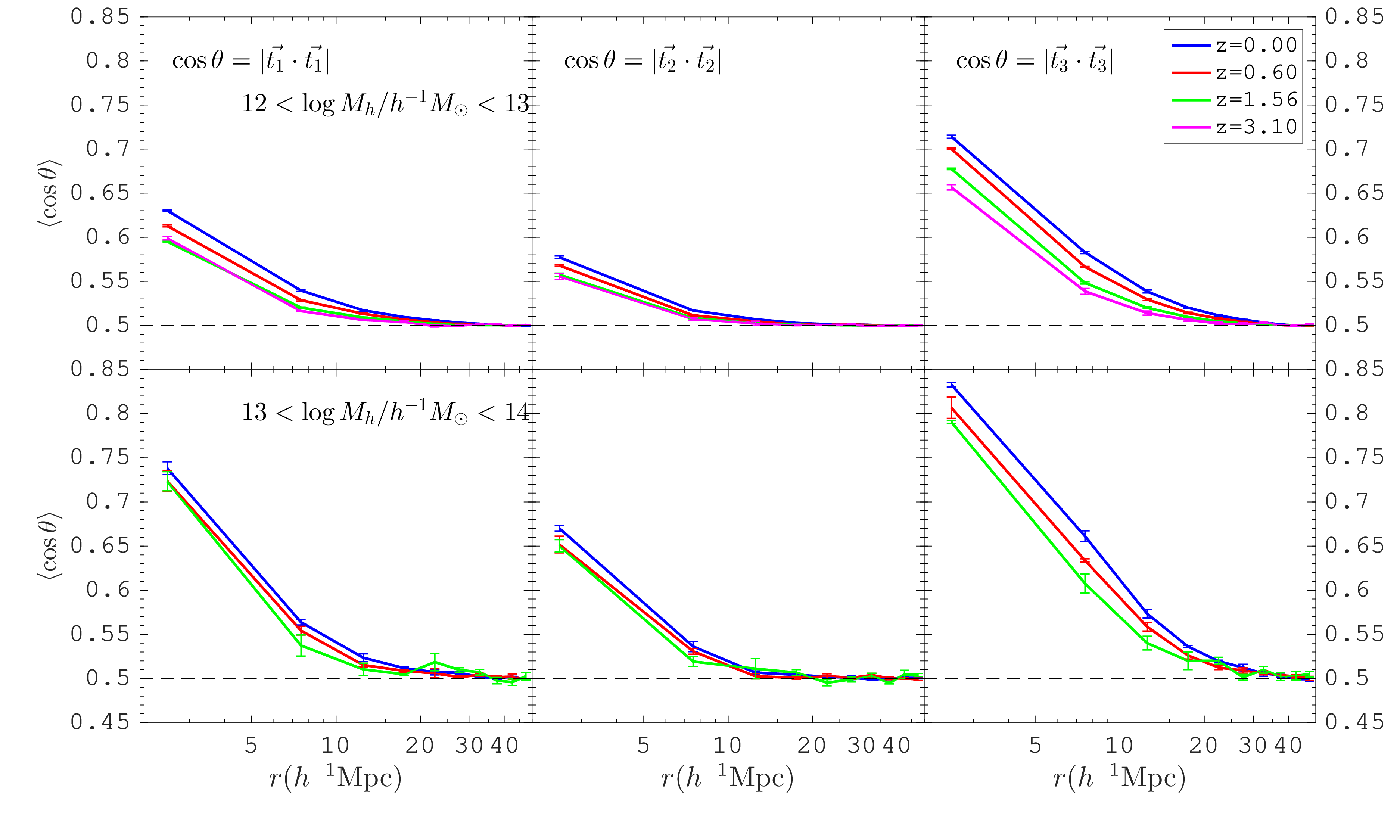}
\caption {Mean alignments between tidal fields on two halos as functions of halo separation for less massive halos(top) and more massive halos(bottom).From left to right: $\cos\theta=|\vt_k\cdot\vt_k|$, $k=1,2,3$.
The colored lines represent the results at different redshift ranges.}\label{fig_ttr}
\end{figure}

We first investigate the tide-tide alignments at the locations of halo pairs
as a function of the pair separation (in co-moving scale).
Fig. \ref{fig_ttr} shows the results at four different redshifts,
which are, respectively, the lowest redshift snapshot in each of the four redshift bins
used above,  and for two halo mass bins, $12\leq\logM<13$ and $13\leq\logM<14$.
In the larger mass bin, the number of halos at $z=3.1$ is too small to
give reliable results, and so the corresponding results are not shown.

There are several interesting trends. (i) The alignment signal decreases with
increasing separation and become marginally important
at distances of $20$ - $30\mpc$. \textbf{This scale may be directly related to
the typical size of large scale structures in the cosmic density field.
Note that the size of our simulation box is only $200\mpc$, which 
may limit the large scale modes we can probe. Thus, the alignment strength
on large scales may be underestimated.} 
(ii) The signal is stronger for massive
halos, which may reflect the fact that more massive halos are
more likely associated with larger structures. (iii) The signal strengthens with
decreasing redshift. This may be due to the fact that large-scale structures
become more prominent as the universe evolves.
(iv) The minor axes of the tidal field are the most strongly aligned,
followed by the major axes and then the intermediate axes.
If the two halos are located within the same large-scale filament, the
major axes of the tidal fields around these halo  are expected to be
aligned because both of the tidal tensors tend to align with
the filaments, as is consistent with our results. However, in this
case it is unclear why the minor axes of the tidal tensors
have the strongest alignment. It may be that most halo pairs on large
scales are not located within the same filamentary structure, but
in two filaments that are embedded in the same sheet-like structure.
Since the minor axes of the tidal fields are perpendicular to the sheet
plane, strong alignments in the minor axes  can be produced.
For the same reason, the alignments of the major axes may be
weakened by cross pairs between two filaments. Thus, our results
may reflect the consequence of the dynamic nature of the cosmic
web, in which halos are embedded in filaments which, in turn,
are embedded in sheets.

\begin{figure}
\plotone{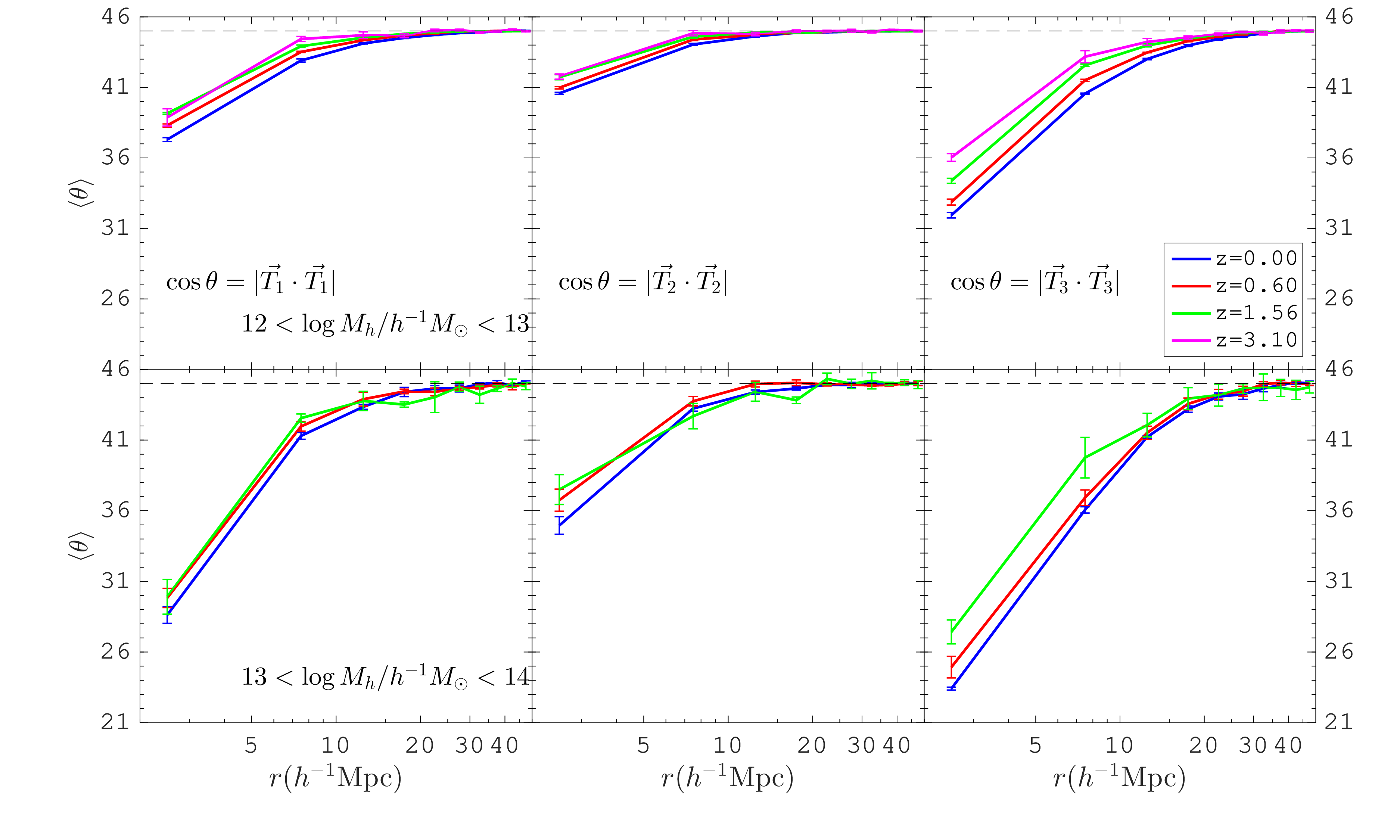}
\caption {Same as Fig. \ref{fig_ttr} but taking into account projection effect. }\label{fig_pttr}
\end{figure}

To make connections to observations, we show in Fig. \ref{fig_pttr}
the alignments of $\vT_k$, the projections of tidal tensors
at the locations of halo pairs. Here we see
again that the alignments can extend to very large scales.
Using the group catalog of Yang et al.  (2007), Lim et al.
(2016, in preparation) have estimated the two-dimensional
tide-tide alignments as a function of separations between galaxy groups,
and found results that are very similar to what we find here.
For example, the mean alignment angle of $\vT_1$ ($\vT_3$) for
groups of $\logM\geq12.5$ is about $35^{\circ}$ ($30^{\circ}$)
at a separation of $3\mpc$,  and approaches $\sim 45^{\circ}$
at $>20 \mpc$, in good agreement with our results. The details of the
comparison between our model predictions and observational
results are presented in Lim et al. (2016).

\subsection{Halo-halo alignments on large scales}
\label{sec_cho}

\begin{figure}
\plotone{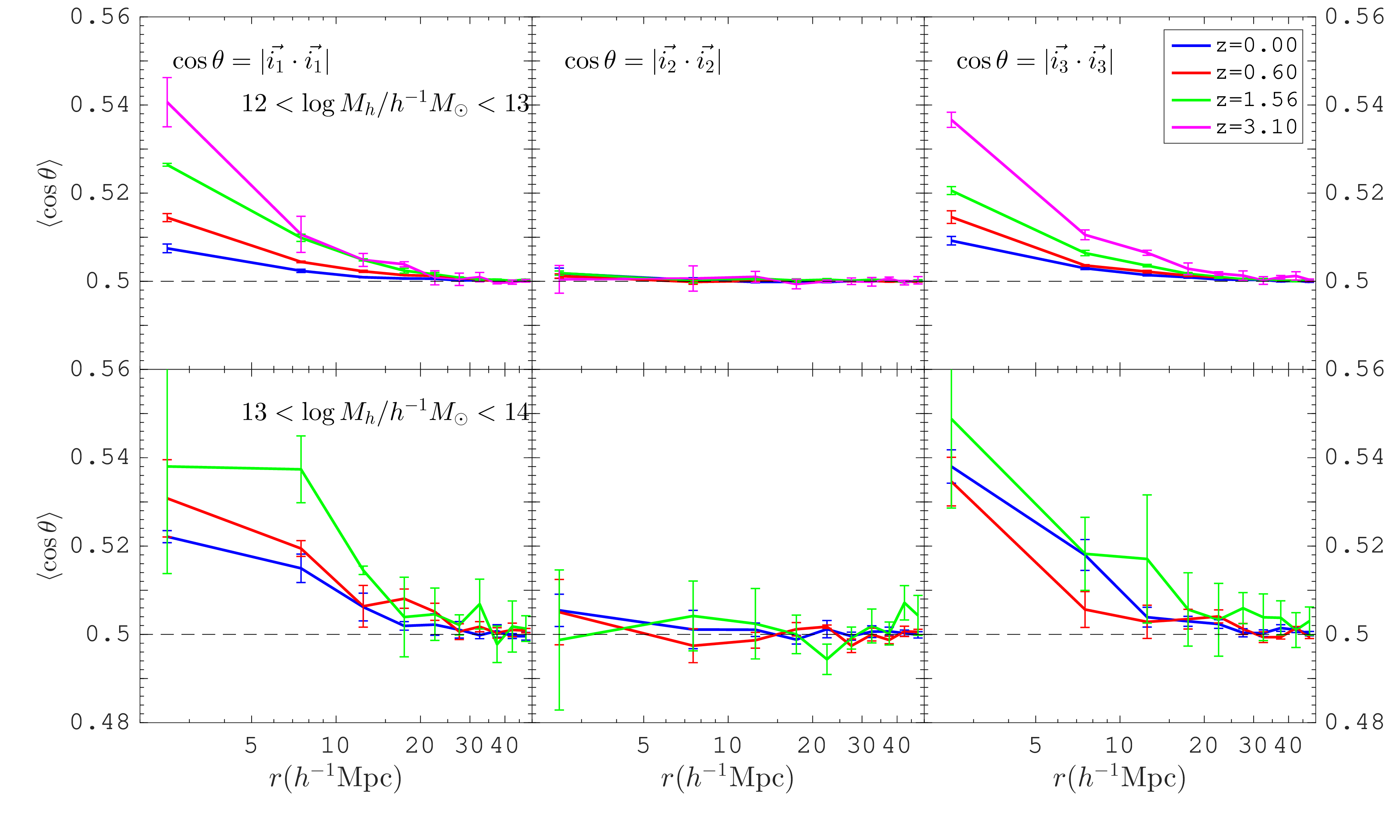}
\caption {Mean alignments between principle axes of two halos as functions of halo separation for less massive halos(top) and more massive halos(bottom).From left to right: $\cos\theta=|\vi_k\cdot\vi_k|$, $k=1,2,3$.
The colored lines represent the results at different redshift ranges.
}\label{fig_iir}
\end{figure}

Fig. \ref{fig_iir} shows the halo-halo alignment as a function of halo pair
separation. Significant alignments are seen only for $\vi_1$ and $\vi_3$.
The alignments are stronger on smaller scales,
vanishing at separations of  $10$ - $20\mpc$. The alignments are
also stronger at higher redshift and for more massive halos.
Lee et al. (2008) measured the ellipticity correlation function and found the
same dependence on redshift and halo mass as we find here
(see also Hopkins et al. 2005).

\begin{figure}
\plotone{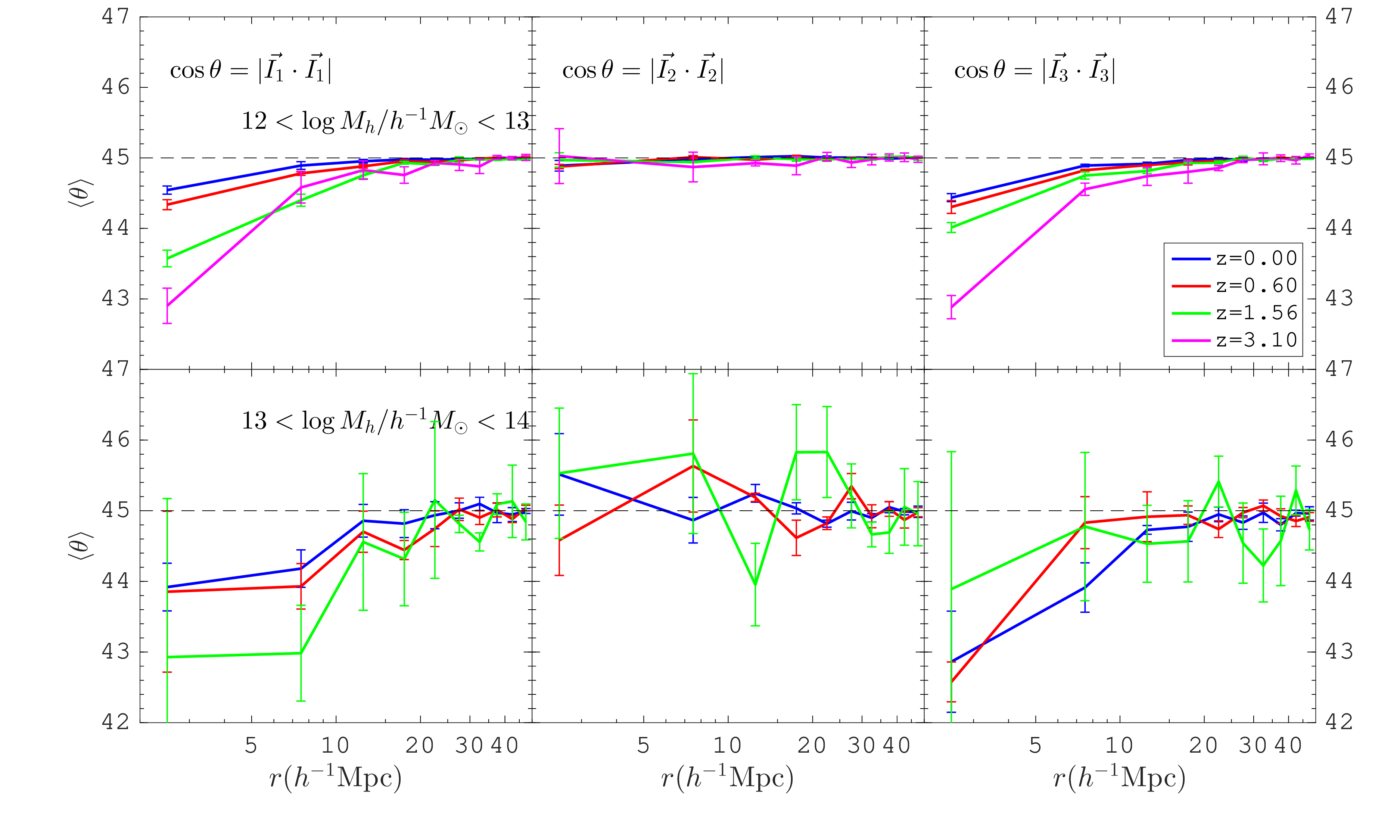}
\caption {Same as Fig. \ref{fig_iir} but taking into account projection effect. }\label{fig_piir}
\end{figure}

In order to facilitate comparison with observation, we also present the
two dimensional results in Fig. \ref{fig_piir}. For massive halos at $z=0$,
the mean angles at $\sim3\mpc$ are $44^{\circ}\pm0.4$ for
$\vI_1$ and $43^{\circ}\pm0.8$ for $\vI_3$.  At higher redshift, the results are
much noisier because of the much smaller number of halos that can be used.
For low mass halos, the mean angle at $\sim3\mpc$ is
about $44.5^{\circ}$, with high significance for both $\vI_1$ and $\vI_3$.
Significant alignments can be seen at least to
$\sim10\mpc$. At $z\sim3$, the mean angles at the smallest scale can
reach $43^{\circ}$.

It is interesting to compare the halo-halo alignments with the tide-tide alignments
and the halo-tide alignments obtained above.
First, the dependence of the halo-halo alignment on halo mass and separation
is very similar to that of the tide-tide alignment, but the
strength of halo-halo alignment is much weaker than the corresponding tide-tide
alignment. Second, the alignments of the major and minor axes of halos
have similar strength. This is in contrast to the tide-tide alignment,
which is the strongest for minor axis, but similar to the halo-tide alignment.
Third, the halo-halo alignment for the intermediate is absent, which
is different from both the tide-tide and halo-tide alignments.
This may be due to the rather weak alignment between the intermediate
axes of halo and tidal field. Finally, the halo-halo alignment increases
with increasing redshift, in contrary to the tide-tide alignment.
However, this is in agreement with the halo-tide alignment, which strengthens with
increasing redshift for a given halo mass. All these together
suggest that the halo-halo alignments on large scales
are produced by  the alignments of halos with local tidal
fields combined with tide-tide alignments on large scales, 
\textbf{with the latter being produced by the large-scale 
coherent structures in the cosmic density field.}

\section{Summary}
\label{sec_sum}

The spin and orientation of galaxies and dark matter halos are found
to be aligned with the cosmic web. Such alignments are important for the
interpretations of gravitational weak lensing observations, as well as
for understanding the formation of galaxies in the cosmic density field.
In this paper, we investigate in detail how various alignments
of dark matter halos depend on redshift and halo mass,
using simulated halos with masses above $10^{12}\msun$
in the redshift range of $\logz\leq0.8$. We use the large-scale
tidal field, estimated from the halo population, to characterize the cosmic web.
The tidal field tensors at halo locations are diagonalized to obtain
the corresponding eigenvectors, $\vt_1$, $\vt_2$ and $\vt_3$ (major, intermediate
and minor axes), with $\vt_1$ corresponding to the stretching direction
of the tidal force, and $\vt_3$ the compressing direction.

We find that the major, intermediate and minor axes
($\vi_1$, $\vi_2$ and $\vi_3$) of halos are aligned with
$\vt_1$, $\vt_2$ and $\vt_3$, respectively. In particular, all the three
alignments generally strengthen with increasing halo mass and redshift.
There are also significant differences among the three alignments.
The halo mass dependence for major axis is stronger at
lower redshift but absent at high redshift, while the results for the
other two axes are almost independent of redshift.

We also investigate the alignment of halo spin ($\vj$) with the local tidal field,
and find that the spin axis tends to be parallel with $\vt_2$ and
perpendicular to $\vt_1$, but the alignment with minor axis is weak.
The strengths of the alignment with $\vt_2$ and anti-alignment
with $\vt_1$ both increase with halo mass and redshift.

We find that once alignments are analyzed for halos of
different peak heights,  $\nu \equiv \frac{\delta_c}{\sigma(M_{h},z)}$,
the dependence on redshift in both the
$\vi_k$-$\vt_k (k=1,2,3)$ and $\vj$-$\vt_2$ alignments disappear,
suggesting that the dependence on halo mass and redshift
is only through $\nu$. We provide accurate fitting formulae
to describe the distributions of the cosine of the alignment angles
as functions of $\nu$. The scaling relations with $\nu$
for the four alignments, $\vi_1$-$\vt_1$; $\vi_2$-$\vt_2$; $\vi_3$-$\vt_3$ and
$\vj$-$\vt_2$, exhibit a similar two-phase behavior, in that the
alignment first strengthens with increasing $\nu$ and then
remains roughly at a constant strength above a transition scale of $\nu$.
We suggest that this is due to the fact that halo formation preserves
the alignment between halo propers and the large scale tidal field in the
linear field, as long as the the large-scale structures remain
in the quasi-linear regime, and that non-linear evolution tends to
suppress the alignment. This scenario also explains why the transition
scales for the three axes are different,  being the largest for minor
axis, along which nonlinear effects start to operate earlier,
and the smallest for major axis, along which nonlinear effects
are the least important.

In order to facilitate comparisons with observations,  we also
investigate the alignments taking into account projection effect.
The overall trends are similar to those in the three dimensional
results, except that the strengths of the alignments are reduced by
projection.

Finally, we investigate the origin of the halo-halo alignments on large scales.
We find that the orientations of the tidal tensors are correlated on scales
up to about $30\mpc$. This, together with the alignments of halos with
local tidal tensors,  implies that halo-halo alignments should also
extend to large scales. Our direct measurements of the halo-halo
alignments confirm this, and the halo mass and redshift dependencies
of the halo-halo alignments can be explained by similar dependencies
in the tide-tide alignment and/or in the halo-tide alignment.

Our results demonstrate that the large-scale tidal field produced by the
large-scale mass distribution in the universe plays a key role in generating
the various alignments observed in numerical simulations. Since the
large-scale tidal field can now be reconstructed from large redshift surveys
of galaxies (e.g. Wang et al. 2012; Wang et al. 2014a). Our results can,
therefore, be used to understand and model the alignments of galaxies and
galaxy systems in the cosmic web. We will come back to this in a future
paper.

\acknowledgments
This work is supported by 973 program (2015CB857005), NSFC
(11522324,11421303),  the Strategic Priority Research Program "The
Emergence of Cosmological Structures" of the Chinese
Academy of Sciences, grant No. XDB09010400 and the Fundamental
Research Funds for the Central Universities. S.J.C is supported by Fund for Fostering Talents in Basic Science of the National Natural Science Foundation of China NO.J1310021. H.J.M.
would like to acknowledge the support of NSF AST-1517528. The numerical
calculations have been done on the supercomputing system in the Supercomputing
Center of University of Science and Technology of China.

\newpage%
\clearpage

\newpage%
\clearpage

\newpage%
\clearpage

\newpage%
\clearpage

\newpage
\clearpage

\newpage
\clearpage

\newpage
\clearpage

\newpage
\clearpage

\newpage
\clearpage

\newpage
\clearpage

\newpage
\clearpage

\newpage
\clearpage

\end{document}